\documentclass[11pt]{article}

\input epsf.sty

\usepackage{graphicx,amsmath,amssymb}
\def\lromn#1{\uppercase\expandafter{\romannumeral#1}}

\begin{document}

\begin{center}
\begin{Large}
\textbf{
Quantum dots as a probe of fundamental physics:
Deviation from exponential decay law
}
\end{Large}

\vspace{1cm}

A. Yoshimi, M. Tanaka,  and M. Yoshimura

Research Institute for Interdisciplinary Science,
Okayama University \\
Tsushima-naka 3-1-1 Kita-ku Okayama
700-8530 Japan

\vspace{7cm}

{\bf ABSTRACT}

\end{center}

\vspace{1cm}
We explore a possibility of measuring deviation from
the exponential decay law in pure quantum systems.
The  power  law behavior 
at late times of decay time profile is predicted in quantum mechanics,
and has been experimentally attempted to detect, but 
with failures except a claim in an open system.
It is found that electron tunneling from resonance state confined in
man-made atoms, quantum dots, has a good chance of detecting the
deviation and testing theoretical predictions.
How initial unstable state is prepared influences greatly
the time profile of decay law, and this can be used
to set the onset time of the power law at earlier times.
Comparison with similar process of nuclear
alpha decay to discover the deviation
is discussed, to explain why there exists a difficulty in this case.

\vspace{4cm}

Keywords
\hspace{0.5cm} 
Exponential decay law and its deviation,
Time evolution of resonance  electron tunneling,
Quantum dot,
Semi-classical approximation,
Nuclear $\alpha-$decay,
Test of quantum mechanical laws,
Exact wave function in parabolic potential

\newpage

\section
 {\bf Introduction}

It is well known that quantum mechanics predicts change of time
profile from the exponential to a power law $\propto t^{-p}\,, p> 0$ 
for times much larger than lifetime defined by the exponential law \cite{peres}.
The power $p$ may differ in various decay processes \cite{power at late}.
To the best of our knowledge this prediction has never been experimentally
confirmed in a definitive way \cite{exp search for power law 1}, 
\cite{exp search for power law 2}.
An exception is a claimed discovery of power law decay
in an open quantum system;
organic molecules dissolved in solvents which have
large relaxation rates caused by interaction
of molecules with environment \cite{rothe et al}.
It is  desirable to verify the  deviation in pure quantum systems
that can neglect interactions with environment.
We shall be able to present presumably the best way
to verify this prediction by using quantum dots,
man-made atoms. A great merit of man-made atoms is
that their parameters are better controllable than natural atoms,
giving good and easy chances of detecting the deviation
at times when remnant atoms are still abundant.
We shall compare this method to similar process of
nuclear $\alpha-$decay.

The power law at late times is  understood as deviation
from the classical Markovian behavior of stochastic process
 given by the exponential decay in which information at the present 
time determines all aspects of the future information.
The Markovian behavior  in  the exponential decay is manifest as 
population ratio at initial time to late times:
the ratio of non-decay probability $P(t)  = | \langle i | e^{- i Ht}|i\rangle |^2$ 
at a later time $t$ to an earlier time $t_0$ is a function of time difference;
$P(t)/P(t_0) = \exp[-(t - t_0)/\tau]$ with $\tau$ the lifetime in
the exponential decay period.
On the other hand, the  power law satisfies an equation,
$\ln \left( P(t_1)/P(t_2)\right) = - p \ln \left( (t_1 - t_p)/(t_2 - t_p) \right)$
in which $t_p\, (\gg \tau)$ is an estimated transition time from the exponential to
the power law.
This formula would reflect that quantum mechanics remembers what
happened at time prior to $t_p$.
If one  finds a non-Markovian behavior  quite different from
this formula predicted by the power law, this might suggest
a breakdown of quantum mechanical law.
In this sense experimental verification of the power law
may be regarded as a test of quantum mechanics.
At earliest times of decay process the non-decay probability
$P(t) $ should behave as an even function
of time,
$1- O(t^2)$ due to hermiticity of hamiltoninan operator $H$ \cite{1-t^2}
unlike the one $1-  t/\tau$ that the exponential decay law predicts.
From  technical difficulty of determining the onset time of decay process
experimental study of earliest time behavior  is more challenging.
We would say somewhat ironically, 
that the intermediate time behavior of exponential decay
law becomes a kind of golden rule, which however must be
modified both at earliest and late times.

\begin{figure*}[htbp]
 \begin{center}
 \centerline{\includegraphics{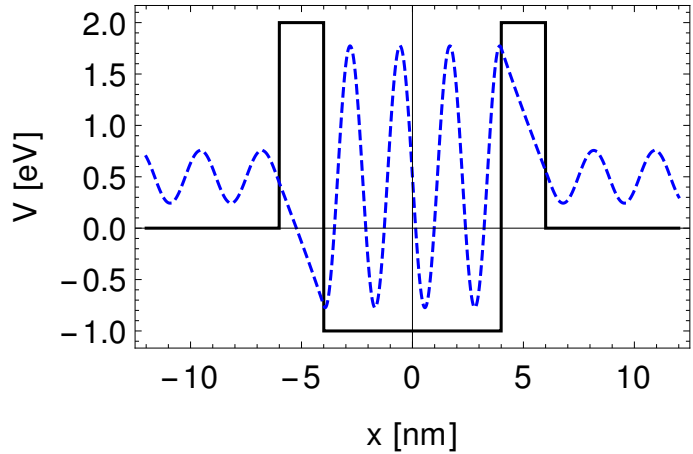}} \hspace*{\fill}
   \caption{
Simplest example of  electron tunneling in
a potential well surrounded by two barriers to create resonant states
whose wave function is illustrated in dashed blue.
}
   \label {resonant well}
 \end{center} 
\end{figure*}

The main reason we use quantum dots as a tool of
exploring fundamental physics is controllability of atomic parameters,
basically atomic size and potential depth \cite{quantum dot}.
Atomic level spacing and their electric dipole moments
in natural atoms are tightly correlated by dominant nuclear Coulomb force,
but in man-made atoms they  are determined by a
human choice of optimized dot size and depth (and barrier height) of potential well.
We shall
consider the size in the range $1 \sim 100 $ nm, the potential depth
and the barrier height in the range $0.1 \sim 50 $ eV.
Basic properties of atoms are worked out in one-dimensional quantum well
and  results may readily be extended to three-dimensional quantum dots.
Typical well structures we consider are illustrated in Fig(\ref{resonant well})
for conceptual one and Fig(\ref{potential shapes}) for more realistic ones.
We shall consider quantum tunneling from a resonant state of electron confined
within surrounding barrier.
The situation is similar to alpha  ($^4$He) particle  tunneling in unstable nucleus
through the Coulomb barrier, but electron resonance energy and its wave function 
in man-made atoms are better calculable.

This paper is organized as follows.
In Section 2 we lay out description of the space-time evolution
of electron tunneling using the semi-classical approximation.
The exponential decay law, along with determination of tunneling
resonance energy and decay width, is derived by imposing an appropriate boundary
condition.
In the next Section 3 we examine how the power decay law is derived
in a class of semi-realistic  potential models
and derive transition time $t_p$ from the exponential to the power law.
It is possible to predict potential parameter dependence of transition time.
In ideal circumstances the transition time may become
$O(10)$ times lifetime of the exponential decay law.
In Section 5 we examine how preparation of electron resonance
influences the time profile of decay law.
Both continuous wave and pulse laser irradiation changes
the time profile, and we clarify how this can be used to make it
easier detection of deviation from the exponential decay.
In Appendix A we present exact solution of the Schroedinger problem
for parabolic potential, which helps to clarify subtle points
in the semi-classical approximation.
Appendix B discusses a problem for potential having cusp structure.
Appendix C explains
the late time behavior of nuclear alpha decay
extending the Gamow potential often used in textbooks.

In a series of works we develop experimental principles of table-top atomic experiments
for fundamental physics using quantum dots.
In another work  \cite{pv paper}  we investigate a possibility  of measuring
parity violation in potential between electrons.

Throughout the present work
we use the natural unit of $\hbar = c = 1$ unless otherwise stated.

\section
{\bf Space-time evolution of electron tunneling viewed as
resonance decay}

We consider one dimensional potential well surrounded by
two barrier walls:
its simplest example is illustrated in Fig(\ref{resonant well}).
It may be more realistic to think of smoothed out potential as in
Fig(\ref{potential shapes}).
One can consider the problem thus set  as an idealization
of more complicated two dimensional well in which
one dimensional electron motion is frozen in the lowest energy level.

The formula we are based on is the semi-classical time dependent wave function
\cite{merz} given by 
\begin{eqnarray}
&&
\psi (x, t) = \int dE f(E) \frac{ \exp[- i E t]}{ \sqrt{ k(x)}}
\left( c_+(E) \exp[ i W(x; E) ] + c_-(E) \exp[- i W(x; E) ] 
\right)
\,,
\label {t-dependent wf}
\\ &&
W(x; E) = \int_0^x dy\, k(y)
\,, \hspace{0.5cm}
k(y) = \sqrt{ 2m \left( E - V(y) \right)}
\,.
\end{eqnarray}
We imagine a wave packet of weight factor $f(E)$
centered around a resonance energy $E=E_*$.
The formula  is appropriate for
 the wave function inside the well, and $x=0$ is taken as
a center of  potential  symmetric under $x \leftrightarrow -x $.
The asymmetric potential can also be considered,
but for simplicity we consider the symmetric case, since
the asymmetric case does not give any new insight into physics issues.
$c_{\pm}(E)$ terms give right-moving and left-moving waves,
respectively, and parity eigenstates dictate $c_{\pm}(E)= \pm c_{\mp}(E)$
for coefficients of parity quantum numbers $\pm$.
From a historical reason the semi-classical approximation
is also called the WKB approximation, or the WKB method.

The wave function in the well is usually connected to 
wave functions under  barriers and to those
outside barriers, using the linearized potential near the classical turning points
given by the equation, $V(x) = E$, since the semi-classical approximation
breaks down at turning points.
There are four turning points
which are denoted by $\pm a(E) \,, \pm b(E)$ with $b(E) > a(E) > 0$.
The linearized potential  near these turning points is parametrized as
\begin{eqnarray}
&&
V(x) - E = - V_a' (a - x)\,, \; {\rm near\;} x = a
\,, \hspace{0.5cm} 
= - V_b'( x-b)\,,  \; {\rm near\;} x = b
\,,
\end{eqnarray}
both of $V_{a,b}'$ being positive.

Time independent Schroedinger equation in the linear potential can
be exactly solved in terms of well-known Airy functions, 
denoted by $A_i(x)$ and $B_i(x)$,
and connection formulas are often derived by using these exact solutions.
We quote connection formulas inside the well and outside two barriers
\cite{merz}:
\begin{eqnarray}
&&
\psi (x; E) =
\frac{A}{\sqrt{k_0(x)} }\exp[i \int_{b}^x k_0(y) dy]  
+   \frac{B}{\sqrt{k_0(x)}} \exp[- i \int_{b}^x k_0(y) dy]\,, 
\; -a < x < a
\\ &&
%\hspace*{-1cm}
\leftrightarrow
%\hspace{0.3cm}
\frac{F}{\sqrt{k(x)}} \exp[i \int_{a}^x k(y) dy]  
+ \frac{G}{\sqrt{k(x)}} \exp[- i \int_{a}^x k(y) dy] \,, 
\; x > b
%\nonumber 
\\ &&
\leftrightarrow
%\hspace{0.3cm}
\frac{J}{\sqrt{k(x)} } \exp[- i \int_{-a}^x dy k(y)] + 
\frac{H}{\sqrt{k(x)}} \exp[ i \int_{-a}^x dy k(y)]\,,
\; x < - b
\,,
\\ &&
\hspace*{-1.5cm}
\left(
\begin{array}{c}
A  \\
 B  
\end{array}
\right)
= \frac{1}{2}
\left(
\begin{array}{cc}
2\theta + \frac{1}{2\theta }  &  - i \left( 2\theta - \frac{1}{2\theta } \right) \\
- i \left( 2\theta - \frac{1}{2\theta } \right)  & 2\theta + \frac{1}{2\theta }   
\end{array}
\right)
\left(
\begin{array}{c}
F  \\
G   
\end{array}
\right)
= \frac{1}{2}
\left(
\begin{array}{cc}
(2\theta+ \frac{1}{2\theta } )\,e^{iL}
 &  - i \left( 2\theta - \frac{1}{2\theta }   \right)\,e^{iL} \\
- i \left( 2\theta - \frac{1}{2\theta} \right)\,e^{- iL}  
& (2\theta + \frac{1}{2\theta })\,e^{- iL}   
\end{array}
\right)
\left(
\begin{array}{c}
H  \\
J   
\end{array}
\right)
\,, 
\nonumber \\ &&
\\ &&
\theta = \exp[\int_a^b dy\, \sqrt{2m (V(y) - E)} ] 
\,, \hspace{0.5cm}
L = \int_{-a}^a dy \sqrt{2m ( E- V(y)\,)}
\,.
\end{eqnarray}
As a means of resonance production we consider a pulsed or CW (continuous wave)
laser irradiation which can excite a bound electron confined in the well  
to a resonance state.
Resonance state is characterized by no incoming wave boundary condition
from outside regions.
This boundary condition implies that $ G=0$ and $H=0 $, which gives 
two conditions,
\begin{eqnarray}
&&
(4 \theta^2 + \frac{1}{4 \theta^2 }) \cos L - 2 i  \sin L = 0
\,,
\\ &&
\frac{F}{J} = - i (4 \theta^2- \frac{1}{4 \theta^2 }) \cos L 
+ 2 \,\frac{ 4 \theta^2 - \frac{1}{4 \theta^2 }}
{4 \theta^2 + \frac{1}{4 \theta^2 } } \sin L \approx 2 \sin L
\,.
\end{eqnarray}
The first equation coincides with the equation that determines
the complex resonance energy $E_*- i \Gamma_*/2$ where
the real part $E_*$ is given by the Bohr-Sommerfield condition of
$L = (n + 1/2) \pi$ (the same semi-classical condition
for bound state energies). For a rectangular potential well
\begin{eqnarray}
&&
E_* = \frac{\pi^2}{8 m a^2} (n + \frac{1}{2})^2 + V_0
\,,
\\ &&
\Gamma_* = \frac{ 1} { 4 m b^2 \theta^2}
\,, \hspace{0.5cm}
\theta = \exp[ \int_a^b dy \kappa(y) ]
\,, \hspace{0.5cm}
\kappa(y) = \sqrt{2m \left( V(y) - E_* \right)} = \sqrt{2m \left( V_1 - E_* \right)}
\,.
\end{eqnarray}
We assumed that the penetration factor $\theta $ is very large,
hence the decay rate is suppressed by $1/\theta^2$.

Exponential decay law follows in a straightforward manner
by considering  barrier penetration after resonance excitation.
A useful approximation for this purpose is to insert the complex resonance energy,
$E = E_* - i \Gamma_*/2$ in the exponent
of eq.(\ref{t-dependent wf}) extended to the region at a
far right by the no incoming wave condition $c_- (E)=0$, to derive
(with a normalization of $F= 1/\sqrt{2\pi}$)
\begin{eqnarray}
&&
\psi ( x,t; E_*) \sim \frac{ f(E_*) c_+( E_*) \Delta E }{\sqrt{2\pi} 
\sqrt{k_*} \theta}  
e^{ i k_* ( x-b) - i E_* t}
\exp[- \frac{\Gamma_*}{2} \left(t - \frac{x- b}{v_*} \right) ]\,
\theta (t - \frac{x- b}{v_*}  )
\,,
\\ &&
| \psi( x,t; E_*)|^2 \simeq
\frac{ |f(E_*) c_+( E_*) |^2 \Delta E^2}{k_*}  4m b^2 \Gamma
\exp[ -\Gamma_* \left(t - \frac{x- b}{v_*} \right)]\,
\theta (t - \frac{x- b}{v_*}  )
\,,
\label {exponential decay of e tunneling}
\end{eqnarray}
where $v_* = \sqrt{2E_*/m}$ is the electron velocity at resonance and
 the wave packet width is assumed to satisfy $\Delta E \gg \Gamma_*$, but much
less than the resonance spacing.

It is however well known that the exponential decay is not the whole
story of decay laws \cite{peres}, \cite{power at late}.
We shall clarify how late time deviation from the exponential decay
arises in potential models, using the space-time evolution
of resonance electron tunneling.

\section
{\bf Potential model and decay law at late times}

\begin{figure*}[htbp]
 \begin{center}
 \epsfxsize=1.0\textwidth
 \centerline{\epsfbox{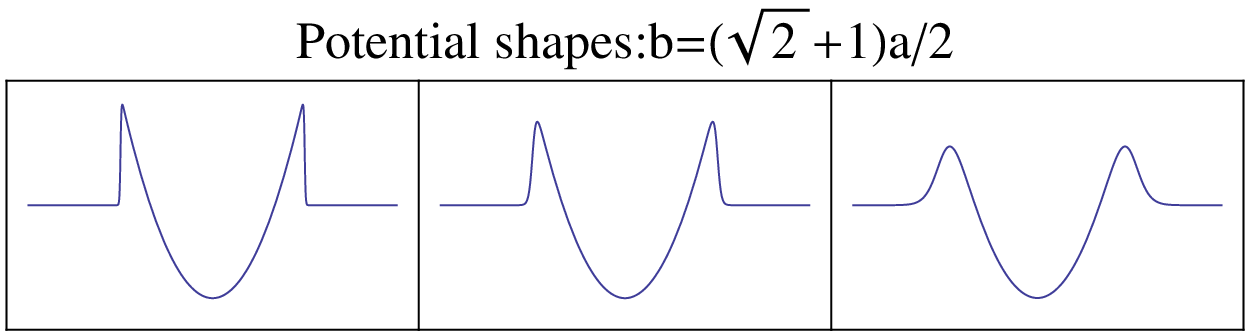}} \hspace*{\fill}\vspace*{1cm}
   \caption{
Illustration of potential shapes for $b/a=(\sqrt{2}+1)/2$:
$(a, \Delta) =$ (5,\,0.5) nm in the left, (5,\,1) nm in the middle, 
and (5,\,2) nm in the right.
}
   \label {potential shapes}
 \end{center} 
\end{figure*}

We consider the following type of potentials characterized by
three size parameters $a\,, b\,, \Delta$ and an overall potential
value $V_w$.
\begin{eqnarray}
&&
V(x) = - V_w \frac{1 - (x^2+a^2)/b^2 }{ 1 + \exp[ (x^2 - a^2)/\Delta^2]}
\,, \hspace{0.5cm}
\sqrt{2} a > b > a
\,, \hspace{0.5cm}
V_w > 0
\,.
\label {general potential}
\end{eqnarray}
The inequality on sizes, $1 < b/a < \sqrt{2}$, is necessary to provide surrounding 
barriers.
Adopting the middle point value of this inequality,
we take $b/a=(\sqrt{2}+1)/2$.
Even with this restriction there are a variety of potential shapes,
as illustrated  in Fig(\ref{potential shapes}).

\begin{figure*}[htbp]
 \begin{center}
 \epsfxsize=0.6\textwidth
 \centerline{\epsfbox{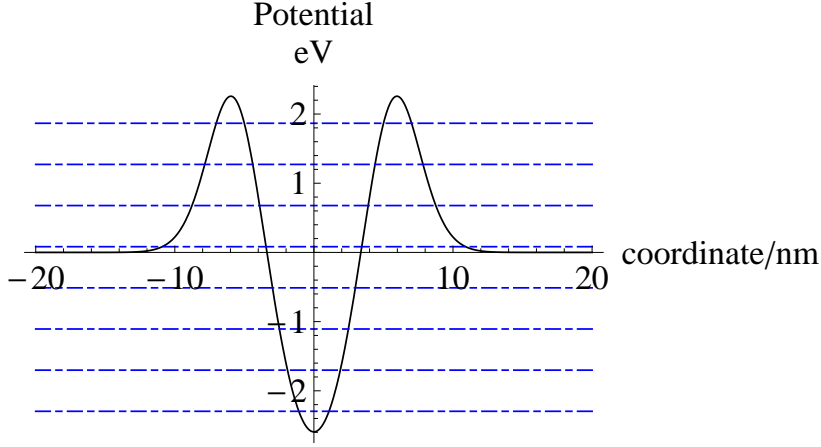}} \hspace*{\fill}\vspace*{1cm}
   \caption{
Potential $V(x)$ based on the formula, eq.(\ref{general potential}), taking
a parameter set:
$a= 5 {\rm nm}\,, b = (\sqrt{2}+1) a/2\,, \Delta =4 {\rm nm}\,, V_w = 10 {\rm eV} $.
Potential maximum is  2.2573 eV, while potential minimum is - 2.5935 eV.
All resonances  and bound state levels  are shown in dash-dotted blue.
}
   \label {potential 2}
 \end{center} 
\end{figure*}

This type of potentials consist of three important parts:
(1) central well part around $x=0$, (2) two barrier regions around
$x = \pm x_{{\rm max}}\,, x_{{\rm max}} > 0$, (3) outside regions
of zero potential. 
The central part is approximately described by harmonic oscillator (HO)
potential given by
\begin{eqnarray}
&&
V_{c}(x) = - V_w ( C_0 - C_2 \, x^2)
\,, \hspace{0.5cm}
C_0  = \frac{ b^2 - a^2}{ b^2 (1 + e^{-a^2/\Delta^2} )}
\,, \hspace{0.5cm}
C_2 = 2e^{a^2/\Delta^2}\,\frac{b^2 - a^2 + \Delta^2 (1 + e^{a^2/\Delta^2} ) }{b^2 \Delta^2 (1 + e^{a^2/\Delta^2} )^2 }
\,.
\end{eqnarray}
Energy levels of HO are given by equally spaced eigenvalues,
\begin{eqnarray}
&&
E_n = \omega_{0}
(n + \frac{1}{2}) -    C_0\, V_w
\,, \hspace{0.5cm}
\omega_{0} = \frac{2}{b} \sqrt{ \frac{2 V_w}{m}} \frac{ \sqrt{ e^{a^2/\Delta^2}
(\frac{b^2 - a^2}{\Delta^2}
+ 1 +  e^{a^2/\Delta^2} })} {1 +  e^{a^2/\Delta^2} }
\,.
\end{eqnarray}
The same formula can also be applied to resonances of energies $E_*$ by using
the semi-classical formula of Bohr-Sommerfield condition,
\begin{eqnarray}
&&
2\,\int_{-c}^c d x \sqrt{2 m \left( E_* - V_{c}(x) \right) } = (n + \frac{1}{2}) \pi
\,,
\end{eqnarray}
with $\pm c $ tuning points given by $  V_{c}(\pm c)= E_*  $.
For an example given by parameters of Fig(\ref{potential 2}), we find 
$E_n = 0.6025 {\rm eV} ( n + 1/2 ) - 2.5935 {\rm eV} \,, \omega_0=0.6025 {\rm eV}$, 
taking $m= 0.1 m_e$.
A useful parameter
$n_0$ may be introduced by a fractional number at which the bound state energy 
vanishes:
\begin{eqnarray}
&&
E_n = \omega_0\, ( n - n_0) 
\,, \hspace{0.5cm}
n_0 = \frac{C_0\, V_w }{\omega_0} - \frac{1}{2}
\,. 
\end{eqnarray}
$n> n_0$ for resonance energy $E_* = E_n$.

Decay widths of resonances are determined 
 by applying the boundary condition of no incoming wave from the outside,
to derive the decay width of n-th resonance ($n> n_0$),
\begin{eqnarray}
&&
\Gamma_n = \frac{\omega_0}{\pi}\theta^{-2}
\,, \hspace{0.5cm}
\theta^{-2} =
\exp[- 4 \pi  (\frac{m}{| V''(x_{{\rm max}} )|})^{1/2} \left( V_{{\rm max}} 
- E_n \right)]
\,,
\label {decay rate formula}
\\ &&
\hspace*{-1cm}
\Gamma_n \simeq 4.00 \times 10^{15} {\rm sec}^{-1} 
\frac{{\rm nm}}{b}\sqrt{\frac{V_w}{ 10 {\rm eV}}}
\exp[ - 4.56 \left(\frac{10 {\rm eV nm}^{-2}}{| V''(x_{{\rm max}} )|} \right)^{1/2} 
\frac{ V_{{\rm max}}  - E_n }{ {\rm eV}} ]\,
\frac{ \sqrt{ e^{a^2/\Delta^2} (\frac{ b^2 - a^2}{\Delta^2} + 1 +  e^{a^2/\Delta^2} )}} {1 +  e^{a^2/\Delta^2} }
\,.
\nonumber \\ &&
\end{eqnarray}
The formula of decay rate, eq.(\ref{decay rate formula}), has a simple interpretation:
by writing the angular frequency $\omega_0 = 2/T_0$ in terms
of classical oscillation period $T_0$ moving within the well,
electron has tunneling probability $\theta^{-2}$ each time electron
arrives at either of two turning points.
In specific examples we consider below
the lifetime given by $1/\Gamma_*$ ranges widely,
of order psec to nsec, or even wider.

The barrier part near potential tops is well described by
parabolic potential, or inverted harmonic oscillator (IVHO):
\begin{eqnarray}
&&
V_b^{\pm}(x) =V_{{\rm max}} - \frac{| V''(x_{{\rm max}}) | }{2} (x \mp x_{{\rm max}})^2
\,.
\end{eqnarray}
Without loss of generality we take the right part of potential at $x \geq 0$,
for which two classical turning points are given by
$V_b^{\pm}(x_{\pm}) = E, \, x_+ > x_- $ for motion of energy $E=k^2/2m$.
Exact solutions of stationary (namely, time independent) Schroedinger equation
for parabolic potentials 
are given by Weber's functions, which are described in Appendix C.
In the present section we shall give results of the semi-classical 
approximation modified by a phase term adopting a feature of
exact result.

We  apply the semi-classical connection formula of complex coordinate
method \cite{ll qm} to wave functions for states near a discrete resonance,
applicable for the parabolic potential:
\begin{eqnarray}
&&
\psi_{b}(x; E) =  (m |V''(x_{{\rm max}} )|)^{-1/8}\,
\frac{e^{-\pi a_u^2} }{(u^2 - a_u^2)^{1/4} }\exp[ \frac{i}{2}
\left( u \sqrt{u^2- a_u^2} + a_u^2 \ln u^2 \right) ]
\,, \hspace{0.5cm}
\\ &&
u = (m |V''(x_{{\rm max}} )|)^{1/4} (x - x_{{\rm max}})
\,, \hspace{0.5cm}
a_u^2 = 2 (\frac{m}{| V''(x_{{\rm max}} )|})^{1/2} ( V_{{\rm max}} - E)
\,.
\end{eqnarray}
The barrier penetration factor $1/\theta^2$ is related to the exponential here by
$\theta^{-2} = e^{- 2\pi a_u^2(E)}$.
This formula is applicable only in a positive $x$ region of
limited $x -x_{{\rm max}} $ range.
As explained in Appendix A, the phase term proportional to $\ln u^2$ is added
by comparing with exact solution of parabolic potential.
Remarkably, every detail in the following can be worked out from this formula.

The space-time evolution at late times is described by
a time dependent wave function;
introducing a wave packet weight $f(E)$, it is
\begin{eqnarray}
&&
\Psi_{b}(x, t ) = 
\int dE f(E) \psi_{b}(x; E) e^{- i E t}
\equiv \int dE f(E) \exp[ i \Phi_b(E; x, t) ]
\,,
\\ &&
\Phi_b(E; x, t) =
\frac{1}{2} u \sqrt{u^2- a_u^2} + i \pi a_u^2 + \frac{i}{4} \ln (u^2 - a_u^2) - Et
\,.
\end{eqnarray}
Note energy dependence via $a_u^2(E)$.
One can estimate this energy integral by the saddle point method.
The exponent function $\Phi_b(E; x,t)$ has a saddle point $E_s$ at
\begin{eqnarray}
&&
\left( \frac{ \partial \Phi_b(E; x,t)}{\partial E}\right)_{E=E_s} = 0
\; \Rightarrow \;
\nonumber \\ &&
E_s \simeq V_{{\rm max}} - \frac{| V''(x_{{\rm max}}) | }{2} (x -x_{{\rm max}})^2
+ \frac{m}{8} \left( \frac{x - x_{{\rm max}}}{t} \right)^2
\,.
\end{eqnarray}
The first two terms in $E_s$ give the parabolic approximation of potential $V_b(x)$
around its top.
The last term is numerically small, since
$\approx 7 \times 10^{-8} {\rm eV}  (x -x_{{\rm max}})^2/{\rm nm})^2
({\rm ps}/t)^2 $, but its presence is important to late time behavior.
When $x$ is close to the turning point and the last term is neglected,
the saddle $E_s$ coincides with resonance energy $E_*$.

The gaussian width around  $E_s$
and the amplitude at the saddle are given by
\begin{eqnarray}
&&
\frac{1}{2} \left( \frac{\partial^2 \Phi_b }{ \partial E^2} \right)_{E=E_s} \approx
\frac{ m}{ 2 |V''(x_{{\rm max}}) |} u \left( u^2 - a_u^2 (E_s) \right)^{-3/2}
= \frac{4}{m} \frac{t^3}{(x -|x_{{\rm max}}|)^2 }
\,,
\\ &&
{\rm with}\hspace{0.3cm} 
u^2 - a_u^2 (E_s) \simeq \frac{m}{4} \sqrt{ \frac{m }{ |V''(x_{{\rm max}}) |}} 
\left( \frac{x - x_{{\rm max}}}{t} \right)^2
\,,
\\ &&
i \Phi_b(E_s) = i W_0(E_s) -
2 \pi \sqrt{\frac{m }{ | V''(x_{{\rm max}})|}}(V_{{\rm max}} - E_s  )
\,, \hspace{0.5cm}
 W_0(E_s) = \frac{u}{2} \sqrt{u^2 - a_u^2(E_s)} -  E_s t
\,.
\end{eqnarray}
The gaussian integral around the saddle gives  amplitude
and probability for the power law decay,
\begin{eqnarray}
&&
\Psi_{{\rm power}}(x, t ) \simeq 
e^{- i \pi/4} \sqrt{\frac{ \pi}{2 }}
 f(E_s)  \frac{ (x - |x_{{\rm max}}| )^{1/2} }{t}
\exp[ i W_0(E_s)
- 2 \pi \sqrt{\frac{m }{ | V_{3}''(x_{{\rm max}})|}}(V_{{\rm max}} - E_s  ) ]
\,,
\\ &&
| \Psi_{{\rm power}}(x, t; E_* )|^2 \simeq \frac{\pi }{2}|f(E_s)|^2 \,
\frac{ x- x_{{\rm max} } } {t^2 }\,
\exp[ - 4 \pi \sqrt{\frac{m }{ | V''(x_{{\rm max}})|}}(V_{{\rm max}} - E_s  ) ]
\,,
\label {late time probability}
\\ &&
\exp[ - 4 \pi \sqrt{\frac{m }{ | V''(x_{{\rm max}})|}}(V_{{\rm max}} - E_s  ) ]
\simeq 
\exp[ - 4 \pi \sqrt{\frac{m }{ | V''(x_{{\rm max}})|}}(V_{{\rm max}} - E_*  ) ]
= \theta^{-2}
\,.
\end{eqnarray}

This amplitude at late times should be compared to
that of the exponential decay period,
\begin{eqnarray}
&&
\hspace*{-0.5cm}
\Psi_{{\rm exp}}( x,t; E_*) \sim \frac{ f(E_*) \Delta E }{\sqrt{2} 
\sqrt{k_*} }  
e^{ i k_* ( x-b) - i E_* t}\,
\sqrt{\frac{\Gamma_* }{ \omega_0}}
\exp[- \frac{\Gamma_* t}{2}  ]
\,,
\end{eqnarray}
with $k_* = \sqrt{2m E_*}$ and $x \ll t$.
Compared to the decay probability at late times,
the exponential decay period gives a corresponding formula,
\begin{eqnarray}
&&
|\Psi_{{\rm exp}}(x,t)|^2  \simeq
|f(E_*)|^2 (\Delta E)^2\, 
\frac{ \Gamma_*}{2 \sqrt{2m E_*} \,\omega_0 } \exp[ - \Gamma_* t]\,
\,.
\end{eqnarray}

Equating two decay probabilities, $|\Psi_{{\rm power}}|^2$
and $|\Psi_{{\rm exp}}|^2$, we derive equation for transition time $t_p$
\begin{eqnarray}
&&
\frac{e^{\Gamma_* t_p} }{ ( \Gamma_* t_p)^2}
=
\frac{| f(E_*)|^2 }{ | f(E_s)|^2}
\frac{ (\Delta E)^2 }{ \Gamma_* k_* \omega_0 ( x- x_{{\rm max} } )   }
\exp[  4 \pi \sqrt{\frac{m }{ | V''(x_{{\rm max}})|}}(V_{{\rm max}} - E_*  ) ]
\,.
\label {transition time eq}
\end{eqnarray}

Let us incorporate the wave packet factor.
We consider pulse laser excitation scheme
from filled bound state electron to a resonant state.
The pulse laser width is denoted by $\Delta \omega = 2\pi \Delta \nu$, 
which may or may not be larger than resonance decay rate $\Gamma_*$.
The wave packet factor is defined by weighted state density
 prepared by laser excitation, hence it is given by
\begin{eqnarray}
&&
f(E) = \frac{(2\pi \Delta \nu)^2/4 }{(E - E_*)^2 + (2\pi \Delta \nu)^2/4 + \Gamma_*^2/4}
\,, \hspace{0.5cm}
\Delta E = 2\pi \Delta \nu = \Delta \omega
\,.
\end{eqnarray}
This leads to
\begin{eqnarray}
&&
\frac{| f(E_*)|^2  }{ | f(E_s)|^2  } (\Delta E)^2 = 
(\Delta \omega)^2 \left( 1 + \frac{4 (E_s - E_*)^2 }{ (\Delta \omega)^2 + \Gamma_*^2}
\right)^2
\,.
\label {wave packet width}
\end{eqnarray}
where $2 \pi \Delta \nu$ may be identified as $\Delta E$
of wave packet width in calculation of space-time evolution
of electron resonance decay.
As previously mentioned, one may estimate $E_s - E_*$ by
taking relevant times $t \geq O(1/\Gamma_*)$, to derive
\begin{eqnarray}
&&
E_s - E_* \leq \frac{ m}{8}\, \Gamma_*^2 (x - x_{{\rm max}})^2
\sim 1.1 \times 10^8 {\rm sec}^{-1} (\frac{x - x_{{\rm max}} }{ {\rm nm}})^2 
(\frac{ \Gamma_*}{{\rm ps}^{-1} })^2
\,.
\end{eqnarray}
We may thus conclude that the ratio, eq.(\ref{wave packet width}),
is very close to $(\Delta \omega)^2 $, and certainly is not much larger than
this value.
In the rest of discussion we shall take the ratio to be $(\Delta \omega)^2 $.

Combining eq.(\ref{transition time eq}) and eq.(\ref{wave packet width}),
we derive for $\xi = \Gamma_* t_p$,
\begin{eqnarray}
&&
\xi - 2 \ln \xi = 4 \pi \sqrt{\frac{m }{ | V''(x_{{\rm max}})|}}(V_{{\rm max}} - E_*  )
+ \ln \frac{ (\Delta \omega)^2 }{ \Gamma_* k_* \omega_0 ( x- x_{{\rm max} } )   }
\label {eq for transition time}
\\ &&
\simeq
X - 26.86 + 
\ln \left( \frac{b }{ x- x_{{\rm max}  } }\frac{\Delta \nu }{ {\rm GHz}} 
\sqrt{ \frac{10 {\rm eV} }{V_w } \frac{{\rm eV}}{E_*}}
\right) \equiv Y
\,,
\\ &&
X = 8 \pi \sqrt{\frac{m }{ | V''(x_{{\rm max}})|}}(V_{{\rm max}} - E_*  ) 
\simeq 28.84 
\sqrt{\frac{{\rm eV nm}^{-2} }{ | V''(x_{{\rm max}})|}}
\frac{ V_{{\rm max}} - E_*  }{{\rm eV}}
\,,
\end{eqnarray}
where  $k_* = \sqrt{ 2m E_*}$.
Note the important relation, the resonance decay rate
$\propto e^{-X/2}$.

The transition region from the exponential to the power law periods
may be better described by adding amplitudes in respective regions,
to give decay probability,
\begin{eqnarray}
&&
| \Psi_{{\rm exp}}(x,t) + \Psi_{ {\rm power}}(x, t ) |^2 \simeq
\frac{(f(E_*)\pi\, \Delta \nu )^2 \Gamma_* }{ 2 k_* \omega_0} 
|e^{- \Gamma_* t/2 - i E_* t} + e^{-i\pi/4} \,e^{- i E_s t}
\frac{ e^{-Y/2} }{\Gamma_* t} |^2
\,,
\label {exp+power time-profile}
\end{eqnarray}
with $e^{-Y/2} = 0.02105$ in our example potential.
Around the transition time $t=t_p$,
there exists an oscillation intrinsic to interference given by the angular frequency 
of  $E_* - E_s$, estimated by
\begin{eqnarray}
&&
E_* - E_s \approx \frac{m}{8 (\Gamma_* t)^2} \left(  
\Gamma_* ( x_+ - x_{{\rm max}})
\right)^2 \simeq 1.1\,  {\rm sec}^{-1}
(\frac{  x_+ - x_{{\rm max}}}{ {\rm nm} })^2 (\frac{ 10\, {\rm nsec} }{ t})^2
\,,
\end{eqnarray}
valid for times $t > O(1/\Gamma_*)$. 
In numerical estimates given below one may take this phase difference
nearly vanishing: $E_s \approx E_*$.

We have examined several cases of parameter choices
to numerically compute, taking $\Delta \omega/2\pi=$
1 GHz,  to derive resonance lifetime, $\xi $ and $ e^{-\xi}$ (remnant fraction).
For this purpose we fix, for simplicity,
 $b= (1 + \sqrt{2})a/2$ taking the half-distance of allowed
edges of required inequalities, $1 < b/a < \sqrt{2}$.
Effective electron mass is taken to be $0.1 m_e$.
Numerical procedure for the estimate is to first assume $a\,, \Delta$
in nm unit for dot sizes and potential value $V_w$, and to numerically
compute various quantities such as
$  n_0\,, |V''(x_{{\rm max}} )|\,, X \,, \xi\,, e^{-\xi}$.
We would prefer a range of outcomes from our own judge 
of experimental easiness;
$1/\Gamma_* = {\rm msec} \sim {\rm nsec}$ for lifetimes
and $e^{-\xi}  > 10^{-10} $  for remnant fraction.
An example that clears these conditions is illustrated below.

\vspace{0.5cm}
\hspace*{-1cm}
\begin{tabular}{|c|c|c|c|c|c|c| c| } \hline 
$(a/{\rm nm}, \Delta/{\rm nm}, V_w/{\rm eV})  $
&    $ E_*/{\rm eV}\, (n, n_0)$ & $1/\Gamma_* $
 &  $ |V''(x_{{\rm max}} )|  $/eV nm$^{-2}$  & X( RHS) & $\xi$ &  $e^{-\xi}$   & $\omega_{\gamma}$/eV\\ \hline
$(5 , 4 ,  10) $   &   1.272 (6, 3.86)  & 17 ns  & 0.83   & 15.59 \,(7.72) & 12.8
& $2.7  \times 10^{-6}  $ & 1.784 \\ \hline
\end{tabular}
\vspace{0.5cm}

\begin{figure*}[htbp]
 \begin{center}
 \epsfxsize=0.6\textwidth
 \centerline{\epsfbox{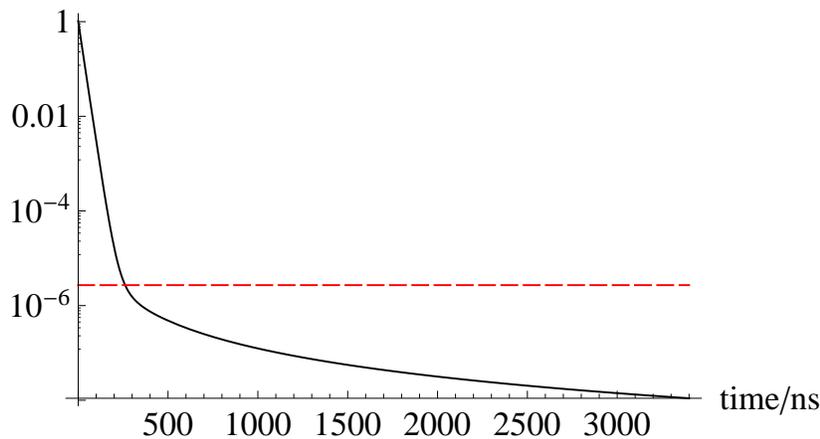}} \hspace*{\fill}\vspace*{1cm}
   \caption{
Time profile of decay law according to
absolute value part of eq.(\ref{exp+power time-profile}).
The horizontal line is population at the transition time.
}
   \label {decay time-profile}
 \end{center} 
\end{figure*}

The  last column shows  laser photon energy from a bound electron of $n=3$
to a resonance state $n=6$.
The potential given by
this parameter set  is depicted in Fig(\ref{potential 2}) along with all discrete levels.
Decay time profile is shown in Fig(\ref{decay time-profile}).

%\begin{{\cal T}*}[htbp]

Our quantum system predicts the late time power to be two,
while results in an open system
\cite{rothe et al} give a variety of late time powers.
It is difficult to provide definite theoretical model for
this open system.

In Appendix B we discuss decay law in partially linear potential similar to the Gamow potential of nuclear $\alpha$ decay and present a problem of late time behavior.
Appendix C presents improved potential model for nuclear $\alpha$ decay.
There is a difficulty of estimating the wave packet weight ratio
in this case.
Moreover, the remaining right hand side factors already give
difficulty of measuring deviation from the exponential decay
in  nuclear $\alpha$ decay.
Much freedom and controllability of electron tunneling in quantum dots
give more promising opportunities  for the new discovery.

\section
{\bf Improved results based on exact numerical solutions of
one dimensional Schroedinger equation}

There are computer softwares to numerically solve the eigenvalue problem
of one-dimensional Schroedinger equation for a given potential.
Our results using the same set of dot parameters in the previous section
are as follows. We listed for comparison results of harmonic approximation, too.

\vspace{0.5cm}
\hspace*{1cm}
\begin{tabular}{|c|c|c|c|} \hline 
$ n$ & $E/{\rm eV} $ & $ \Gamma/{\rm eV} $ & HO $ E(\Gamma)$/eV\\ \hline \hline
$0 $ & $ -2.28  $ & $ 0 $ &  $ -2.296 $\\ \hline
$1 $ & $ -1.67 $ & $0 $ &  $-1.7015   $ \\ \hline
$ 2$ & $-1.06 $ & $0 $ & $-1.107 $ \\ \hline
$3 $ & $ -0.472 $ & $ 0 $ & $-0.512 $ \\ \hline
$ 4$ & $ 0.0909 $ & $ 6.5 \times 10^{-12} $ & $ 0.0827$ \\ \hline
$5 $ & $ 0.640 $ & $ 5.9  \times 10^{-8} $ & $ 0.677 $ \\ \hline
$ 6$ & $ 1.16 $ & $1.3  \times 10^{-5} $ & $ 1.272(1.93 \times 10^{-8})$ \\ \hline
$7 $ & $ 1.64 $ & $9.6  \times 10^{-4} $ & $1.867 $ \\ \hline
$8 $ & $ 2.06  $ & $ 2.5  \times 10^{-2} $ & $2.4615 $ \\ \hline
$ 9$ & $ 2.43 $ & $N/A $ & None \\ \hline
%$ $ & $ $ & $ $ & \\ \hline
\end{tabular}
\vspace{0.5cm}

\begin{figure*}[htbp]
 \begin{center}
 \centerline{\includegraphics{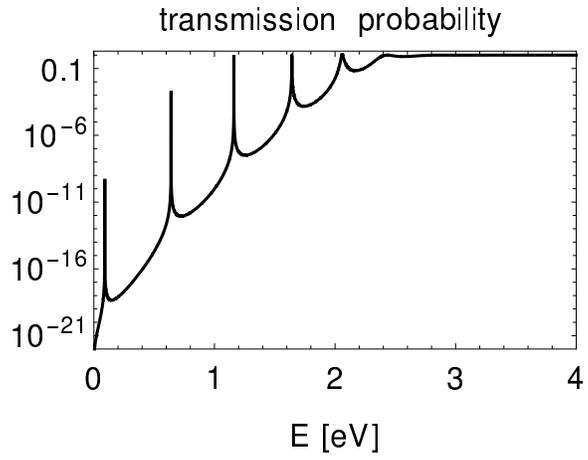}} \hspace*{\fill}
   \caption{
Transmission probability passing double barriers.
}
   \label {transmission prob}
 \end{center} 
\end{figure*}

Resonance energy and their width determination is non-trivial.
We solved the Schroedinger equation for one-way boundary condition
at right and computed reflection and transmission probability at far left.
Energy dependence of this probability shows resonance behaviors
at special energies, as shown in Fig(\ref{transmission prob}).
From these  we determined resonance energy and
their width which give results as in the above table.
We note as a minor comment that transition probability determined this way
corresponds to double barrier penetration, while electron resonance decay
after laser excitation is given by a single barrier penetration.
Hence a factor 1/2 of resonance decay is required for correct interpretation.

The transition time to the power law period 
is estimated by using exact results  as follows.
One calculates $X$ and $Y$ factor using the resonance energy $E_*$,
to derive $X= 34.74$ and $Y=9.60$, from which one has the transition time
and the remnant fraction,
\begin{eqnarray}
&&
t_p = \frac{\xi }{\Gamma_*} = 15.02 \times 5.05 \times 10^{-11} {\rm sec}
= 0.76 \,{\rm ns} %0.759
\,, \hspace{0.5cm}
e^{-\xi} = 3.0 \times 10^{-7}
\,.
\end{eqnarray}

%\begin{{\cal T}*}[htbp]

\begin{figure*}[htbp]
 \begin{center}
 \centerline{\includegraphics{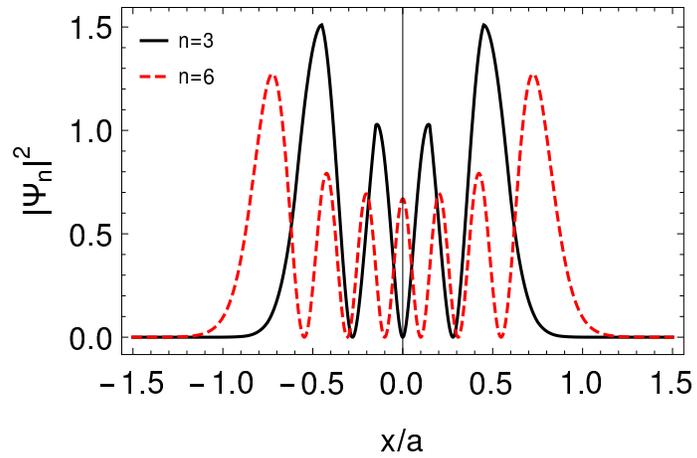}} \hspace*{\fill}
   \caption{
Absolute value of squared
wave functions of $n=6$ resonance  in dashed red and $n=3$ bound  state
in solid black.
}
   \label {wave functions 3-6}
 \end{center} 
\end{figure*}

Discrepancy between exact results and those of harmonic approximation
increases as quantum number increases, because the approximation is based
on truncation of potential to power series at the bottom.
The discrepancy is copious in particular for resonances.
In experimental design one needs radiative transition rate for
$| 6 \rangle \rightarrow |3\rangle$.
Using numerically derived wave functions of resonance and bound state
as shown in Fig(\ref{wave functions 3-6}),
we computed dipole transition amplitude, to obtain
\begin{eqnarray}
&&
|\langle n=3|x|n=6 \rangle | = 0.0124  {\rm nm}
\,. 
\end{eqnarray}
In the harmonic approximation this matrix element vanishes due to
the selection rule $\Delta n = \pm 1$.
The semi-classical approximation, on the other hand, gives
a rather large value 1.15 nm.
We shall use dipole moment given by the numerical computation above.
The dipole transition amplitude and energy difference between these states, $ 1.632$eV
(to be compared with HO result $1.784 $eV),
give radiative transition rate, $2.55 \times 10^5$sec$^{-1}$.

\section
{\bf Decay law under continuous preparation of unstable states}

It was pointed out \cite{optically activated decay},
\cite{fluctuation-assisted decay} that rapid and persistent fluctuations such
continuous wave (CW) excitation to resonance states may lead to modified
late time power decay.
The problem is related to that  measurements
in quantum mechanical systems do depend on how initial states
are prepared.
We shall study this problem using optical Bloch equation
under laser irradiation
in electron resonance state that incorporates effectively the
power law period as a non-constant decay rate.

We introduce three-level optical Bloch equation for pure system
in order to discuss the decay law when resonance state is
prepared by laser irradiation.
We denote three states by $|i \rangle \,, i = 1,2,3$
in which $|1\rangle $ is resonance state, $|2 \rangle $ is bound state,
and $|3 \rangle $ is outgoing plane-wave state in continuum.
In the interaction picture we may write two-level OBE under irradiation,
\begin{eqnarray}
&&
\frac{d\sigma_{11}}{dt} = i \frac{\Omega(t)}{2} 
(\sigma_{12} - \sigma_{21})
-  \Gamma_p(t) \sigma_{11} - \gamma_{12}  \sigma_{11} (1- \sigma_{22})
\,,
\\ &&
\frac{d\sigma_{22}}{dt} = - i \frac{\Omega(t)}{2} 
(\sigma_{12} - \sigma_{21} )
+  \gamma_{12}  \sigma_{11}(1- \sigma_{22})
\,,
\\ &&
\frac{d\sigma_{21}}{dt} = - i \delta_L \sigma_{21}
- i \frac{\Omega(t)}{2} (\sigma_{11} - \sigma_{22})
- \frac{\gamma_{12}  }{2} \sigma_{21}(1- \sigma_{22})
\,,
\\ &&
\frac{d\sigma_{12}}{dt} = i \delta_L \sigma_{12}
+ i \frac{\Omega(t)}{2} (\sigma_{11} - \sigma_{22})
- \frac{\gamma_{12} }{2} \sigma_{12}(1- \sigma_{22})
\,,
\\ &&
\frac{d\sigma_{33}}{dt} = \Gamma_p(t) \sigma_{11}
\,,
\end{eqnarray}
where $\gamma_{12}$ is single-photon decay rate 
$|1 \rangle \rightarrow |2 \rangle + \gamma$, and 
\begin{eqnarray}
&&
\delta_L = \omega_L - \omega_{12}
\,, \hspace{0.5cm}
\Omega(t) = - \langle 2 | \vec{d} | 1 \rangle \cdot \vec{E}(t)
\,.
\end{eqnarray}
Rotating wave approximation is made.
The factor $(1- \sigma_{22}) $ is not usually present, but
we introduced this term as effect of Pauli blocking of already occupied state.
We introduce  amplitude time dependence $E_0(t)$ in 
equation $\vec{E}(t) = E_0(t) e^{i \omega_L \,t + i \Phi(t)}$ of laser field, but
not phase time dependence ($\Phi(t)$ set to zero) for simplicity.

We may choose the power decay formula 
and laser intensity as follows:
\begin{eqnarray}
&&
\Gamma_p(t) = - \frac{d}{dt} \ln\, |e^{-\Gamma_* t/2} + e^{-i\pi/4}\,\pi
\frac{\sqrt{k_*(x- x_{{\rm max}} )} }{2E_* (t +\tau)}  |^2
\,, \hspace{0.5cm}
\tau = \frac{1}{\Gamma_*}
\,,
\\ &&
\Omega (t) = \frac{\Omega_0 (t) }{2} 
\left( 1 - \frac{2}{\pi} {\rm Arctan} \frac{t -T_1 }{\Delta  } \right)
 \,, \hspace{0.5cm}
\Omega_0 (t) =  \langle 2 | d | 1 \rangle \cdot E_0(t)
\,.
\end{eqnarray} 
Laser switch-off time   $T_1 $  may be taken in a wide range,
for instance much less than $\tau$ to of order power law transition
time.
The width $\Delta$ is assumed to be of order the laser width $\approx 1 $GHz.
The factor $\tau$ in the power law amplitude $\propto 1/(t+\tau)$
cannot be precisely fixed, but it should be sufficient for this estimate.
It is important to note that radiative E1 decay of $|1 \rangle \rightarrow |2 \rangle$
is described excellently by the exponential law without the power period.

Since two time scales $\gamma_{12}$ and transition time $O(10)/\Gamma_*$
are vastly different, one may separate optical Bloch equation into time spans of
the laser irradiation and the late time behavior. 
The first part may be solved by setting $\Gamma_p(t) =0$ and taking
a constant CW laser of time independent Rabi frequency $\Omega$.
The steady state solution of $t \rightarrow \infty$ may be derived by
taking vanishing derivative, to give
\begin{eqnarray}
&&
\sigma_{11}  = \frac{\Omega^2}{ 4 (\delta_L^2 + \gamma_{12}^2/4 )}
\,, \hspace{0.5cm}
\sigma_{22} = 1 - \sigma_{11} 
\,,
\\ &&
\sigma_{12} = - \frac{\Omega }{2 (\delta_L - i \gamma_{12}/2) }\, \sigma_{11}
\,.
\end{eqnarray}
By adjusting Rabi frequency and detuning one may keep
a large population of resonance state along with a large coherence with
the bound state.

Two interesting limiting cases are then described as follows.
The first case concerns a short time termination of laser irradiation.
In this case resonance decays both into bound state and outgoing plane wave state.
Our special quantum dot case gives a fast radiative decay into bound state,
and this reduces greatly the possibility of observing the power decay law
of resonance tunneling.

\begin{figure*}[htbp]
 \begin{center}
 \epsfxsize=0.6\textwidth
 \centerline{\epsfbox{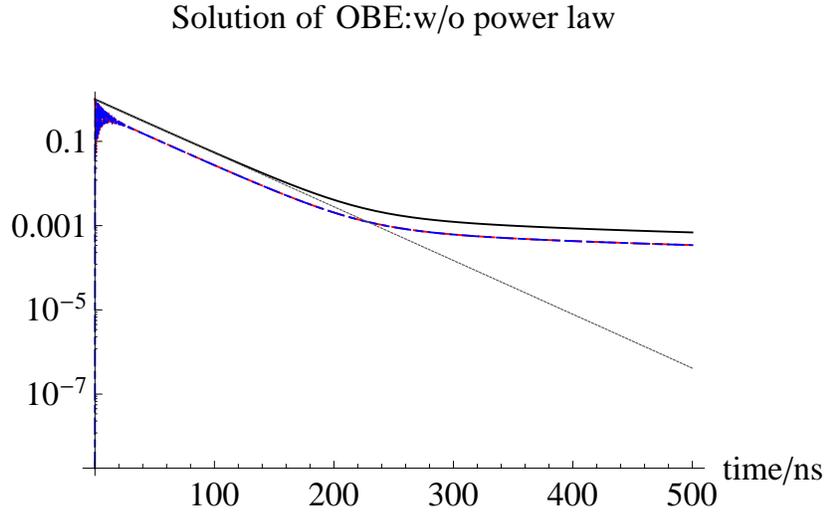}} \hspace*{\fill}\vspace*{1cm}
   \caption{
Time profile of non-decay probability given by
 $1-\rho_{33}(t)$ (with $\rho_{33}=\sigma_{33}$ total population within
quantum dot) under CW laser irradiation of zero detuning:
10 W cm$^{-2}$ in solid black, 1 W cm$^{-2}$ in dashed red,
1 W cm$^{-2}$ without the power law period in dash-dotted blue,
and the case without laser irradiation in dotted black.
}
   \label {decay time-profile under cw}
 \end{center} 
\end{figure*}

The second case concerns CW laser irradiation over the whole range
of power law period.
As illustrated in Fig(\ref{decay time-profile under cw}),
decay process is modified under laser irradiation, without much changing
the transition time to the power law period,
as in the case of Fig(\ref{decay time-profile}).
Quantity adopted here as the non-decay probability is $1 -\sigma_{33}(t)
= \sigma_{11}(t) + \sigma_{22}(t)$,
which contains fast Rabi oscillation between bound state (given by
population $\sigma_{22}(t)$) and resonance.
We thus see that how preparation of resonance states is made
under laser irradiation has a profound effect of controlling the
electron resonance tunneling.

\begin{figure*}[htbp]
 \begin{center}
 \epsfxsize=0.6\textwidth
 \centerline{\epsfbox{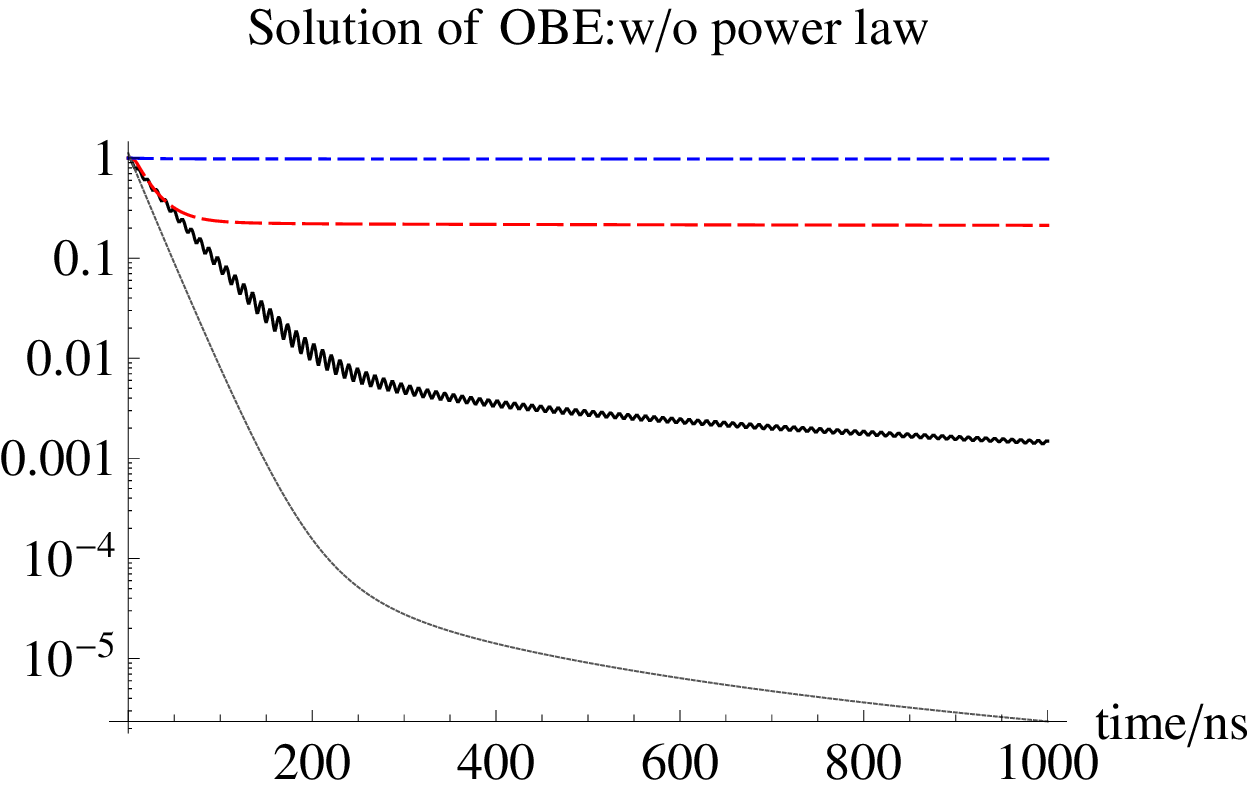}} \hspace*{\fill}\vspace*{1cm}
   \caption{
Time profile of decay law $1-\rho_{33}$
under laser irradiation of zero detuning:
1 W cm$^{-2}$ under CW in solid black, 1 W cm$^{-2}$ under pulse
of duration 10 ns in dashed red,
pulse 1 W cm$^{-2}$ of duration 1 ns in dash-dotted blue,
and the case without laser irradiation in dotted black.
}
   \label {decay time-profile under cw+pulse}
 \end{center} 
\end{figure*}

Duration of pulse laser determines the exponential period as
illustrated in Fig(\ref {decay time-profile under cw+pulse}):
shorter pulse expedites onset time of power law period.
This way one can control to a certain extent
experimental design for discovery of deviation from the exponential decay.

\section
{\bf Summary}

Late time power law has been derived for smooth potentials
of electron tunneling in man-made atoms.
The late time power law $\propto 1/t^2$ was derived and
the transition time $t_p$ is determined in terms of
potential parameters, which gives in favorable conditions
$O(10) \times $ lifetime of the exponential decay.
In man-made atoms, atomic size,  potential depth and barrier height
can be arranged independently, and this made it possible
to derive transition times from the exponential to the power
short enough that the power law may be experimentally observed
with remnant fractions as large as of order $10^{-7}$.
The crucial factor for this arrangement is roughly size $\times$
square root of potential height, a typical combination of
penetration factor against barriers.

How unstable resonance state is prepared influences
the decay law, and we verified this using laser irradiation for
transition from bound state.
Both CW and pulse laser irradiation changes the time profile of
decay law, and duration of pulse influences the onset time
of transition to the power law.
This helps discovery of deviation from the exponential decay.

A similar tunneling decay, nuclear alpha decay, is predicted to
obey the same power law $\propto 1/t^2$, but
has experimental difficulty due to a large transition time from
the exponential to the power law.
We have also examined radiative atomic decay, nuclear beta decay,
and some other decay processes, but in all cases
it is difficult to find a promising possibility of experimentally observing
deviation from the exponential decay law.
Man-made atoms give a unique and interesting opportunity of
studying deviation of the exponential decay at late times.

\section
{\bf Appendix A:  Exact  solutions
of the Schroedinger equation for parabolic potential}

The stationary Schroedinger equation of energy $E$ in a parabolic potential
\begin{eqnarray}
&&
- \frac{1}{2m} \frac{d^2 \psi}{d x^2} + \left( V(x) - E \right) \psi = 0
\,, \hspace{0.5cm}
V_2(x) = V_1 - \frac{|V_0''|}{2} x^2
\,,
\end{eqnarray}
with $V_1 > 0$,
may be transformed to Weber's differential equation,
\begin{eqnarray}
&&
\frac{d^2 w}{d z^2} + (\lambda + \frac{1}{2} - \frac{z^2}{4} ) w = 0
\,,
\end{eqnarray}
by a change of coordinate variable and their parameters,
\begin{eqnarray}
&&
z = \alpha x
\,, \hspace{0.5cm}
\alpha = e^{ i \pi/4} (m |V_0''|)^{1/4}
\,, \hspace{0.5cm}
\lambda = - \frac{1}{2} - 2i (\frac{ m}{|V_0''| })^{1/2} (V_1 - E)
\,.
\end{eqnarray}
Classical turning points for electron moving with energy $E < V_1$ are at
\begin{eqnarray}
&&
\pm a
\,, \hspace{0.5cm}
a = \sqrt{\frac{ 2(V_1 - E)}{|V_0''|}}
\,.
\end{eqnarray}

Two independent solutions of Weber's differential equations are
Weber's functions denoted by $D_{\lambda}(z)$ and $D_{-\lambda - 1}(iz)$
\cite{whittaker-watson}.
These functions can be written in terms of linear combinations of
confluent hypergeometric functions $F(\alpha, \gamma; z^2/2)$
of parameters, $(\alpha, \gamma) = (-\lambda/2, 1/2)\,, (1/2 - \lambda/2, 3/2)$.

Using general formulas of asymptotic expansion for confluent hypergeometric
functions, one confirms that a specific linear combination of two fundamental solutions
gives the boundary condition of no incoming wave at the far right.
This solution at $x>a$ is given by, disregarding normalization,
\begin{eqnarray}
&&
\psi(x; E) =
\frac{\sqrt{2\pi}e^{i\pi/4} }{\Gamma( \frac{1}{2} - i a_u^2) }
e^{- \frac{\pi}{2} a_u^2}\, D_{ - \frac{1}{2} - i a_u^2} ( e^{i \pi/4}\, u )
- 
 D_{ - \frac{1}{2} + i a_u^2} ( e^{ i 3 \pi/4}\, u)
\,,
\end{eqnarray}
and its leading term at $x \gg a$ is
\begin{eqnarray}
&&
%\approx 
\psi_R (x; E) =
\frac{- i }{\left( ( m |V_0''|)^{1/4} x \right)^{1/2} } 
\frac{e^{-\pi a_u^2} }{(u^2 - a_u^2)^{1/4} }\exp[ \frac{i}{2}
\left( u \sqrt{u^2- a_u^2} + a_u^2 \ln u^2 \right) ]
\,, \hspace{0.5cm}
\\ &&
a_u = (mV_0'')^{1/4} a = (\frac{m}{|V_0''|})^{1/4} \sqrt{2 (V_1 - E)}
\,, \hspace{0.5cm}
u = (m |V_0''|)^{1/4} x
\,.
\end{eqnarray}
The formula $e^{-{\cal P}_b}\,,  {\cal P}_b = \pi  ( m| V_0''|)^{1/2} a^2$ 
may be regarded as an effective
barrier penetration factor for the parabolic potential.
On the other hand, the leading term at $x \ll -a $ is given by
\begin{eqnarray}
&&
\frac{2 i e^{i\pi/4} }{\left( ( m |V_0''|)^{1/4} | x|  \right)^{1/2} } 
\cos \left( \frac{1}{2} ( u \sqrt{u^2- a_u^2} + a_u^2 \ln u^2 )
+ \frac{\pi}{4}
\right) 
\,.
\label {wf at left}
\end{eqnarray}

\begin{figure*}[htbp]
 \begin{center}
 \centerline{\includegraphics{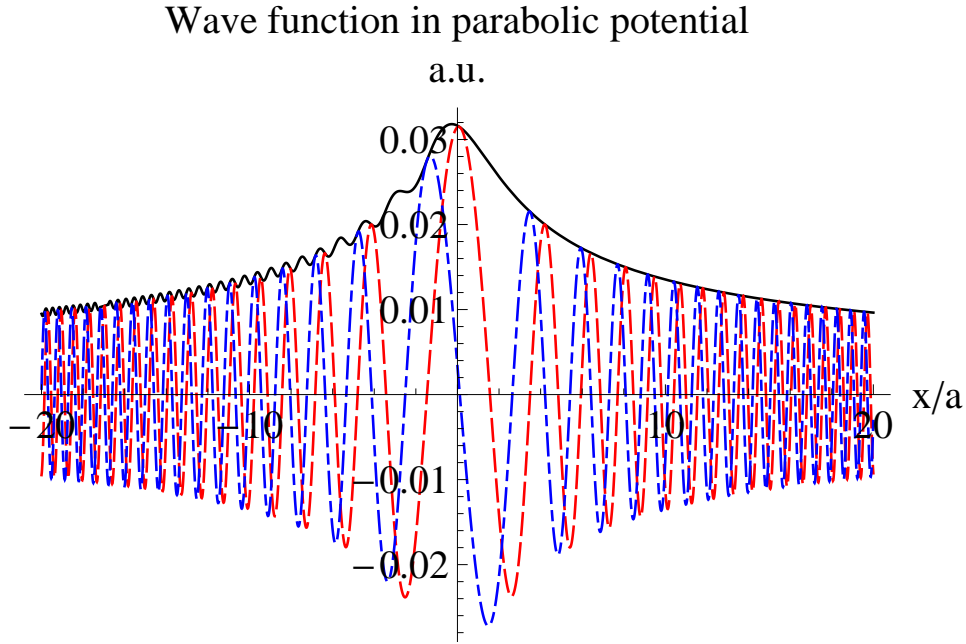}} \hspace*{\fill}
   \caption{ 
Solution given by Weber's function for $a_u = (m |V_0''|)^{1/4} a = 2$:
absolute value in sold black, imaginary part in dashed red,
and real part in dash-dotted blue.
}
   \label {weber sol}
 \end{center} 
\end{figure*}

Using the asymptotic formula of exact solution, one may estimate the probability
at turning points, $x = \pm \sqrt{ 2(V_1-E)/|V_0''|} \equiv \pm a$,
to give
\begin{eqnarray}
&&
%|\psi (\pm a)|^2 \approx 
\frac{1}{ (m |V_0''| )^{1/4} a} \exp[- 2\pi (m |V_0''| )^{1/2} a^2 ]
= \sqrt{\frac{\pi  }{ {\cal P}_b}}  \exp[ - 2 {\cal P}_b]
\,.
\label {finite amp at tp}
\end{eqnarray}
In Fig(\ref{weber sol}) we illustrate exact solution given by
Weber's function, which shows that the left side sinusoidal
wave goes out of a parabolic potential to a  right-moving component alone.
The penetration factor \\
$\exp[- \pi (m |V_0''|)^{1/2} a^2/2 ] = 0.21$
is only modestly small in this example.

\begin{figure*}[htbp]
 \begin{center}
 \centerline{\includegraphics{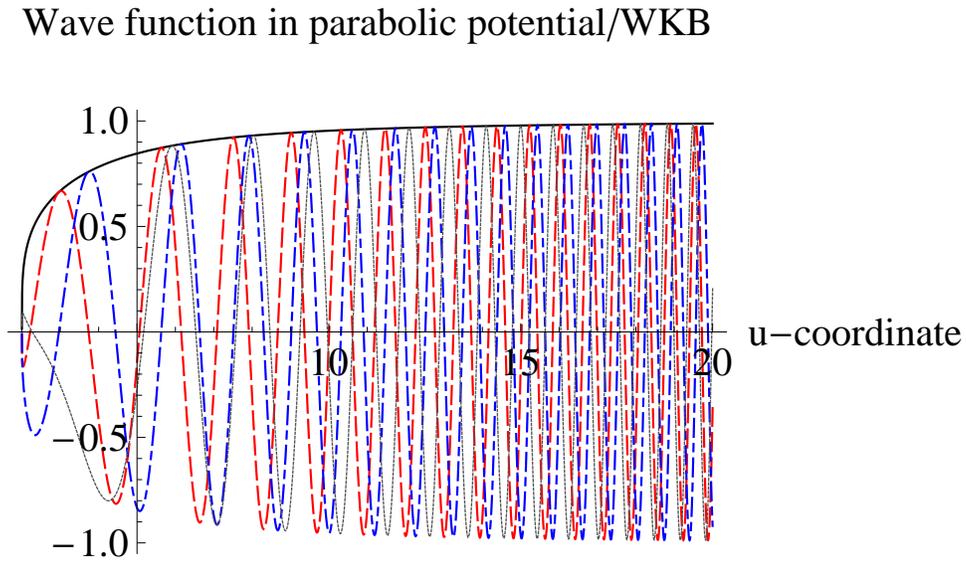}} \hspace*{\fill}
   \caption{ 
Ratio of exact solution to the semi-classical result
in the case of $a_u = (mV_0'')^{1/4} a = 2$:
absolute value in solid black, imaginary part in dashed red,
real part in dash-dotted blue, and the semi-classical result without
$\propto \ln x^2$ phase factor in dotted black.
}
   \label {wkb comparison}
 \end{center} 
\end{figure*}

Although exact solutions are valuable,
its mathematical complexity is often demanding for its full understanding.
The semi-classical or WKB approximation is intuitively appealing,
giving a clear relation to classical behaviors.
The wave functions in classically allowed region of $|x| > a$ 
is described by linear combinations of running waves;
at $x>a$
\begin{eqnarray}
&&
\frac{1}{\sqrt{ 2m \left( E - V_2(x) \right)}}
\exp[\pm i \int_a^x dy\, \sqrt{ 2m \left( E - V_2(x) \right)} ]
\,.
\end{eqnarray}
We impose the boundary condition of no incoming wave from the far right,
which chooses one of these waves.
Connection passing turning points to potential region under the parabolic
barrier and then to the left of $x < -a$
 can  be carried out,
following the method of complex coordinate \cite{ll qm}.
This connection gives  wave function in the far left region
of the form similar to eq.(\ref{wf at left}), 
except the phase term $\propto \ln u^2 \propto \ln x^2$.
In Fig(\ref {wkb comparison}) we compare thus modified semi-classical result
relative to exact result.

%\begin{{\cal T}*}[htbp]

\section
{\bf Appendix B: Problem of partially linear potential}

\begin{figure*}[htbp]
 \begin{center}
 \centerline{\includegraphics{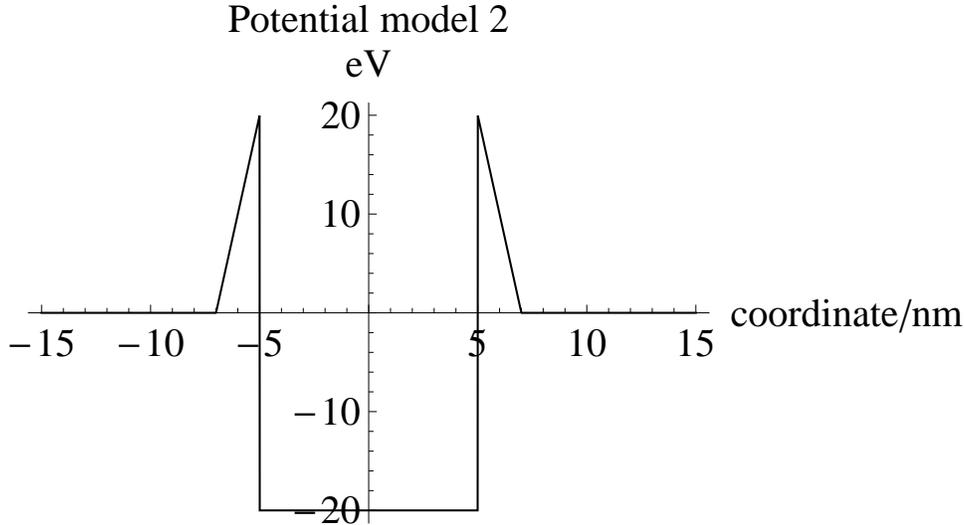}} \hspace*{\fill}
   \caption{
Example of potential given by eq.(\ref{d-lin potential})
for the case of $a = 5$nm, $V_0 = 20\,, V_1= 20$ eV's,
and $V_a' = 20{\rm eV}/ 2{\rm nm}$.
}
   \label {double linear well}
 \end{center} 
\end{figure*}

We  consider a partially linear potential as depicted in Fig(\ref{double linear well}).
It is a $x\leftrightarrow -x $ symmetric potential of the form,
\begin{eqnarray}
&&
\hspace*{-1cm}
V_1(x) = V_1^+(x) + V_1^+(-x)
\,, \hspace{0.5cm}
V_1^+(x) = 
 - V_0 \theta(a - x ) \theta (x)
 + \left( V_1 - |V_a'| (x - a) \right)
\theta ( x-a ) \theta(b- x) 
\theta (x) 
\,,
\label {d-lin potential}
\end{eqnarray}
with $V_0\,, V_1 $ taken positive.
This potential has an aspect of cusp structure
(at $x= \pm a$) common to Gamow potential 
often used in discussion of nuclear $\alpha$ decay.
The cusp in Gamow model occurs at nuclear radius.

Consider the right half $x>0$ of this potential.
Resonance solution is characterized by no incoming wave boundary condition 
from $x>b$, proportional to $e^{i k x}\,, k > 0$.
Such a solution  is given by a linear combination of two Airy functions, $A_i(z)\,,B_i(z)$:
 at $a < x \leq b$:
\begin{eqnarray}
&&
C_i(x) = N \left( A_i (\zeta(x)\,) + i B_i  (\zeta(x)\, ) \right)
\,, \hspace{0.5cm}
\zeta(x) =  (\frac{2}{3} \frac{\sqrt{2m} }{ |V'_a|} )^{2/3}\, \left( E 
- V_1 + |V_a'| (x - a)   \right)
\,,
\end{eqnarray}
with $N$ a normalization constant.
This combination of Airy functions has asymptotic behavior given by
\begin{eqnarray}
&&
C_i(x) \propto x^{-1/4} \exp[i \sqrt{2m |V_a'| }\, x^{3/2} ]
\,.
\end{eqnarray}
As shown in \cite{ll qm}, this wave function has a constant flux
despite of a fast phase variation $\propto x^{3/2} $.

At a far right of $x > b$ the function $C_i(x) $ is connected to
outgoing plane-wave $D(x) $ given by
\begin{eqnarray}
&&
D(x) = C_i(b) \exp[i \sqrt{2m E} (x-b)]
\,.
\end{eqnarray}
Within the well at $x < a$ the wave function is given by
sinusoidal function, either a cosine or sine function 
depending on parity of states, of variable $\sqrt{2m (E+V_0)}\, x $.
Connection of wave functions at turning points $x= a$
requires matching of logarithmic derivatives:
\begin{eqnarray}
&&
\hspace*{-1cm}
\sqrt{2m (E+V_0)}
\left(
- \tan (\sqrt{2m (E+V_0)}\,a ) \,,
\; {\rm or}\; 
\cot (\sqrt{2m (E+V_0)}\,a ) \right)
=   (\frac{2}{3} \frac{\sqrt{2m} }{ |V'_a|} )^{2/3}\,|V'_a|
\left(A_i' (\zeta(a)\,)  + i B_i' (\zeta(a)\,) \right)
\,,
\nonumber \\ &&
\end{eqnarray}
two respective equations
corresponding to  parities of states.
This equation  determines the resonance energy $E= E_*$.

The asymptotic form of Airy functions is adequately described by
semi-classical approximation \cite{ll qm}.
The space-time evolution at late times may thus be investigated by
working out  the semi-classical   wave function:
\begin{eqnarray}
&&
\psi (x, t) = \int dE f(E) \frac{ c_+(E)}{ \sqrt{ k(x)}}
  \exp[ i \Psi(E, x) ] \,, \hspace{0.3cm}
x > b
\,,
\label {t-dependent wf outside}
\\ &&
\Psi(E, x) = W(x; E) -  E t
\,, \hspace{0.5cm}
 W(x; E) = \int_b^x dy\, k(y)
\,, \hspace{0.5cm}
k(y) =
\sqrt{ 2 m \left( E - V_1(y) \right)} 
\,.
\end{eqnarray}
Estimate of the energy integral in this equation may be given by
the saddle point method.
Saddles are found  by vanishing condition of the phase derivative,
\begin{eqnarray}
&&
\left( \frac{\partial \Psi (E, x)}{ \partial E} \right)_{E=E_s}=
\pm \frac{ \sqrt{2m}}{|V'_a| } (E_s - V_1)^{1/2} +
\sqrt{\frac{m}{2E_s}} x - t =0
\,.
\end{eqnarray}
Time  $t$ is taken large at late times,
 and the saddle point energy $E_s$ is found of order, $E_s = O( t^2 |V'_a|^2/m)$.
The gauusian width at the saddle necessary for the energy integral is given by
\begin{eqnarray}
&&
\frac{\partial^2 \Psi}{ \partial E_s^2}
= \pm \frac{1}{2} \frac{ \sqrt{2m}}{|V'_a| } (E_s - V_1)^{-1/2} 
- \frac{1}{4} \frac{ \sqrt{2m} }{ E_s^{1/2}}\,x
\,.
\end{eqnarray}
This gaussian width 
$| \frac{\partial^2 \Psi}{ \partial E_s^2}|^{-1/2}$ %at relevant $E_s$ 
is in proportion to
$ E_s^{1/4} \propto t^{1/2}$  for large $E_s$, which implies,
combined with the pre-factor $1/\sqrt{k}=1/(2m E_s)^{-1/4} $, 
a  constant decay probability $\propto t^0$,
implying that the decay process terminates at latest times after the
exponential period.
We regard this result inappropriate for description of decay processes. 
The result was confirmed by using exact Airy solutions as well.

The possibility of a  constant decay probability after the exponential period 
has been noted in a solvable model \cite{onley-kumar}.
In this particular model the non-decay at latest times is attributed
to existence of bound state which appears in
a strong coupling regime beyond perturbation theory.
Our case of partially linear potential model does not 
correspond to this case of emergent bound state.
It is thus not clear under what conditions the behavior
of constant decay probability at late times emerges.

\section
{\bf Appendix C: Late time behavior  of nuclear $\alpha$ decay 
}

We  clarify   late-time profile of $\alpha$ decay by calculating 
probability amplitude based on  Woods-Saxon attractive 
$\alpha-$nucleus potential  more realistic than
the Gamow potential often used in textbooks of nuclear physics.
A modified potential is also necessary from result of Appendix B
that led to termination of decay process.

\subsection
{\bf $\alpha$-nucleus interaction based on Woods-Saxon potential}

Consider interaction of $\alpha$ particle ($^4$He nucleus) with residual nucleus, consisting
of attractive force inside nucleus and electro static force outside nucleus.
We assume that  proton or charge distribution inside the residual nucleus
has the same form as an attractive Woods-Saxon potential $V_{WS} (r) $,
\begin{eqnarray}
&&
V_{WS} (r) = - V_0 v(r)
\,, \hspace{0.5cm}
 v(r) = \frac{1}{ 1 + \exp[(r-R)/a] }
= \frac{1}{2} ( 1 - \tanh \frac{r-R }{ 2a} )
\\ &&
V_0 = ( 51 - 33 \frac{N-Z }{ A} )\, {\rm MeV}
\,, \hspace{0.5cm}
a = 0.67\, {\rm fm}
\,, \hspace{0.5cm}
R = 1.27 A^{1/3} \, {\rm fm}
\,.
\end{eqnarray}
This potential incorporates a thin nuclear effect
instead of cusp at nuclear radius $R$ in the Gamow potential
$A, Z, N$ are atomic number, atomic nuclear charge and neutron number of nucleus.

\begin{figure*}[htbp]
 \begin{center}
 \centerline{\includegraphics{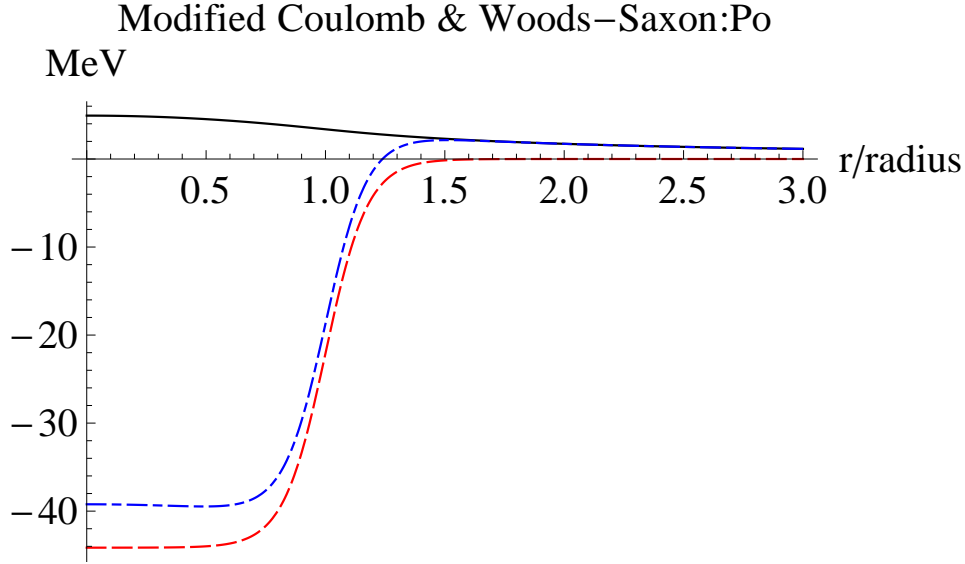}} \hspace*{\fill}
   \caption{ %{\bf To be replaced.}
$\alpha$ decay potential of $^{212}_{84}$Po and its components:
modified Coulomb in sold black, Woods-Saxon attraction in dashed red, and
their sum in dash-dotted blue.
}
   \label {potential components}
 \end{center} 
\end{figure*}

Electro static potential $V_C$ between $\alpha$ and residual nucleus can
be calculated from the Poisson equation,
\begin{eqnarray}
&&
\vec{\nabla}^2 V_C = 2 (Z-2) e^2 N v(r)
\,, \hspace{0.5cm}
N^{-1} = \int d^3 r v(r) = \frac{R^3}{c} 
\,, \hspace{0.5cm} 
\\ &&
\frac{1}{c} = 2\pi \int_0^{\infty} dx \, x^2 \left(
1- \tanh \frac{R}{2a} (x-1)
\right)
= \frac{1}{0.2216} \; {\rm for}\; A = 212
\,,
\\ &&
V_C (r) = \frac{ 2 (Z-2) e^2 }{4\pi} N \int d^3 \rho \frac{v(r) }{ |\vec{r} - \vec{\rho} |}
= \frac{(Z-2) e^2 c}{R^3} \left( \frac{1}{r} \int_0^r dy\, y^2 + \int_r^{\infty} dy\, y
\right) \frac{1}{ 1 + e^{(y-R)/a}}
\,.
\end{eqnarray}
The constant $N$ is determined by the total charge of
$\alpha-$residual nucleus.
The remaining radial integral is expressed in terms of
polylogarithmic functions of n-th order, $Li_n (z)$:
\begin{eqnarray}
&&
V_C (r) = - \frac{ (Z-2) e^2\,c  }{  Li_3 (- e^{R/a} )} \frac{r^2}{a^3} J(r)
\,,
\\ &&
(\frac{r}{a})^2 J(r) = \frac{\pi^2}{6} + \frac{1}{2} 
(\frac{R}{a})^2 - \frac{1}{6} (\frac{r}{a})^2
- Li_2 ( - e^{(r-R)/a} ) 
+ 2 \frac{r}{a} Li_3 ( - e^{(r-R)/a} )  - 2 \frac{r}{a}Li_3 ( - e^{-R/a} ) 
\,.
\end{eqnarray}
The total potential acting on $\alpha$ is given by
\begin{eqnarray}
&&
V_{\alpha} (r) = V_{WS} (r) + V_C (r)
\,.
\end{eqnarray}
Example of $^{212}_{84}$Po is illustrated in Fig(\ref{potential components}).
%and Fig(\ref{potential derivatives}).

\subsection
{\bf Exponential decay law}

The time profile  of $\alpha-$decay
viewed far away from nucleus is derived using
transition amplitude ${\cal T}(E)$, a function of $\alpha$ energy $E$,
\begin{eqnarray}
&&
\psi_{\alpha} (r, t) = \int_{-\infty}^{\infty} dE f(E) {\cal T} (E)  e^{i k(E) ( r -R ) - i Et } 
\,,
\end{eqnarray}
and is dominated by barrier penetration of $\alpha$ particle.
Mechanism of resonant $\alpha$ particle formation
is not well understood, but the time evolution is essentially
dictated by the boundary condition of no incoming wave from
the outside of Coulomb repulsion.
This time interval is described in a similar way to
resonance formation of electronic state in man-made atoms,
and one can assume the formula of eq.(\ref{exponential decay of e tunneling})
adapted to this potential case,
\begin{eqnarray}
&&
\psi_{\alpha} ( x,t; E_*) \sim \frac{ f(E_*) C_+( E_*) \Delta E }{\sqrt{2\pi} 
\sqrt{k_*} \theta}  
e^{ i k_* ( x-b) - i E_* t}
\exp[- \frac{\Gamma}{2} \left(t - \frac{x- b}{v_*} \right) ]
\,,
\\ &&
| \psi_{\alpha} ( x,t; E_*)|^2 \simeq
\frac{ |f(E_*) C_+( E_*) |^2 \Delta E^2}{k_*\, \theta^2} 
\exp[ -\Gamma \left(t - \frac{x- b}{v_*} \right)]
\,, \hspace{0.5cm}
\\ &&
\theta^{-2} =  4M R^2 \Gamma =
\exp[- 2 \int_{-R}^{R} d\rho \sqrt{ 2M \left( V_{\alpha} - E \right)}]
\,,
\end{eqnarray}
with $C_+( E) $ being the amplitude hitting at the barrier.
Details of the potential is irrelevant to
the exponential decay law, and the most important factor of barrier penetration
factor $1/\theta$ is given in terms of imaginary momentum
$\kappa(\rho) = \sqrt{ 2M \left( V_{\alpha}(\rho) - E \right)}$  integral.

\subsection
{\bf Late time behavior in the semi-classical approximation
and estimate of transition time}

Global features of $\alpha$ decay potential are similar to
potential form given in Section 4, except energy (MeV vs eV)
and size (fm vs nm) scales.
The decay probability formula given for electron resonance decay,
eq.(\ref{late time probability}), may be adopted in $\alpha$ decay
by appropriate changes of physical quantities from atoms to nuclei.

Estimate of transition time $T$ using $\xi = \Gamma T$ is thus given by
\begin{eqnarray}
&&
\xi - 2 \ln \xi = \ln \frac{\pi r  }{2 v(V_m) M R^2 }
+ 2 \ln \frac{f(Q)}{ f(V_m)}
\,,
\label {transition time eq alpha}
\\ &&
\frac{f(Q)}{ f(V_m)} \simeq 1 + \frac{ 4 (V_m - Q)^2}{ (\Delta E)^2}
\,.
\end{eqnarray}

We imagine observations of $\alpha$ particle at a fixed distance $r$ from
nuclear targets, taken 1 m in the calculation. 
The largest target number at transition time
 is estimated as $e^{-42.98} = 2.16 \times 10^{-19}$ in this example.

A few important features of eq.(\ref{transition time eq alpha}) are

(1) independence of both $Q$ and $\Gamma$, which suggests that
lifetime of order 1 sec to 1 hour may be of experimental interest.

(2) choice of measurement distance $r$: distance closer to target region
is favored. Additional dependence away from 10 m is $ \ln (r/ 10 {\rm m})$.
At $r= 10 $cm, $\Gamma T =40.56 \,, e^{-\Gamma T} \sim 2.4 \times 10^{-18}$.

(3) atomic mass number dependence $- (2/3)\ln A$, which favors large mass number

Experimental efforts of searching for deviation from
the exponential decay have been most extensively
 focused on nuclear beta and alpha decays.
Record search time is $45 T_{1/2}$ in nuclear beta decay
\cite{exp search for power law 2}.
and $40 T_{1/2}$ in nuclear alpha decay \cite{exp search for power law 1}
without any positive result for the deviation.
There are reasons why the search has met difficulties in these cases.

\vspace{1cm}
 {\bf Acknowledgements}

This research was partially
 supported by Grant-in-Aid 18K03621(MT), 20H00161(AY), and 17H02895 (MY)  
from the
 Ministry of Education, Culture, Sports, Science, and Technology.


\begin{thebibliography}{99}
\bibitem{1-t^2}
The argument for this is as follows:
$| \langle i | e^{- i Ht}|i\rangle |^2 \simeq  | \langle i |(1 - i H t - \frac{H^2}{2} t^2
+ O(t^3) |i\rangle |^2 \simeq 1 - ( \langle i | H^2 |i\rangle - 
\langle i | H |i\rangle^2 )t^2 = 1- t^2\, \sum_{n\neq i} | \langle n | H |i\rangle|^2$.
\end{thebibliography}
\end{document}